\documentclass[
reprint,
superscriptaddress,
 amsmath,amssymb,
 aps,
 prl,
floatfix,
]{revtex4-2}

\usepackage{lipsum}
\usepackage{color}
\usepackage{natbib}
\usepackage{graphicx}% Include figure files
\usepackage{dcolumn}% Align table columns on decimal point
\usepackage{bm}% bold math
\begin{document}

%\preprint{APS/123-QED}

\title{A granular scaling approach to landslide runout}

\author{Rory T. Cerbus}
 \email{rory.cerbus@riken.jp}
%  \thanks{Current affiliation: RIKEN Center for Biosystems Dynamics Research (BDR), 2-2-3 Minatojima-minamimachi, Chuo-ku, Kobe 650-0047, Japan}

\affiliation{RIKEN Center for Biosystems Dynamics Research (BDR), 2-2-3 Minatojima-minamimachi, Chuo-ku, Kobe 650-0047, Japan
}
\affiliation{
Laboratoire Ondes et Mati\`{e}re d'Aquitaine (LOMA), UMR 5798 Université Bordeaux et CNRS, 351 cours de la Lib\'{e}ration, 33405 Talence, France
}
% \homepage{https://www.loma.cnrs.fr/rory-cerbus/}

\author{Ludovic Brivady}
\affiliation{
Laboratoire Ondes et Mati\`{e}re d'Aquitaine (LOMA), UMR 5798 Université Bordeaux et CNRS, 351 cours de la Lib\'{e}ration, 33405 Talence, France
}

\author{Thierry Faug}
\affiliation{
Univ. Grenoble Alpes, CNRS, INRAE, IRD, Grenoble INP, IGE, 38000 Grenoble, France
}

\author{Hamid Kellay}
\affiliation{
Laboratoire Ondes et Mati\`{e}re d'Aquitaine (LOMA), UMR 5798 Université Bordeaux et CNRS, 351 cours de la Lib\'{e}ration, 33405 Talence, France
}


\date{\today}% It is always \today, today,

\begin{abstract}
A main objective in landslide research is to predict how far they will travel. Landslides are complex, and a complete understanding in principle requires accounting for numerous parameters. Here we engender a simplification by investigating the maximum landslide runout using granular laboratory experiments and a scaling analysis. We find that correctly accounting for the fall height and grain size distribution not only yields an improved correlation of normalized runout, but also quantitatively unites laboratory and field data. In particular, we find that the mobility of landslides increases with the square root of the fall height and with the skewness of the grain size distribution.
\end{abstract}

\maketitle

The feature of landslides and avalanches of principal interest to the inhabitants of mountainous regions is how far they will reach. The cost in life and property in affected areas is enormous \cite{keefer2007assessing,perkins2012death}, with ecological hazards sometimes remaining decades later \cite{newhall2000}. Predicting the runout distance, and in particular what parameters control it, is thus a major objective of landslide and avalanche research \cite{legros2002mobility}. The variety and complexity of these parameters has encouraged a varied approach to this problem, and landslides and avalanches with different features are often treated as distinct. Highlighting the importance of features such as the slide geometry or soil content, landslides are commonly subdivided into nearly 30 different categories including debris flows and rock avalanches \cite{hungr2014varnes}. Snow avalanches are likewise treated as an entirely separate species of large mass movement \cite{atwater1954snow}. A confounding problem is the difficulty to isolate and systematically study important parameters in nature. While laboratory experiments using small grains can methodically study features such as longitudinal ridges \cite{magnarini2019longitudinal}, the role of interstitial fluids \cite{iverson1989dynamic,iverson2000acute}, the overrun of dams \cite{faug2008overrun}, the generation of tsunamis \cite{miller2017tsunamis}, or the onset of avalanche behavior \cite{daerr1999two}, these small-scale experiments are often argued to have only limited relevance to large-scale natural landslides and avalanches because of their vastly different time and length scales \cite{iverson2015scaling,davies1999runout,delannay2017granular}. Thus the prevailing consensus, albeit not universal \cite{lucas2014frictional}, is that while laboratory granular flows, snow avalanches, debris flows, and rock avalanches all share common features, their runout cannot be comprehended using a single approach.

Although an uncomplicated description of landslide runout appears to be out of the question, simple trends have been known for decades \cite{legros2002mobility}. A classical result common among all natural systems is that the runout distance increases if the landslide is larger or falls from a greater height \cite{legros2002mobility,johnson2017drop,delannay2017granular,lucas2014frictional}, although the exact dependence is not known. (For the sake of brevity we will hereafter refer to all natural and laboratory mass movements collectively as ``landslides", regardless of their specific, intricate details.) This correlation between runout distance $L$, fall height $H$, and landslide volume $V$ is typically demonstrated through plots of $H/L$ versus $V$, where the non-dimensional ratio $\mu = H/L$ is sometimes termed the effective friction or Heim's ratio \cite{heim1932bergsturz}. Although the scatter in these plots can be substantial \cite{legros2002mobility}, it is generally agreed that as $V$ increases, $H/L$ decreases. Such plots thus demonstrate the possibility of uncovering a profound simplification in an otherwise impossibly complicated system, even if this is widely considered an incomplete description of landslide behavior \cite{davies1999runout,johnson2017drop,legros2002mobility,lucas2014frictional}. %The downward trend in $H/L$ with increasing $V$ does not obviate the dependence of $L$ on other landslide parameters, but it does suggest that $L$ depends predominantly on $H$ and $V$. 

The present work builds on the classical result represented by traditional plots of $H/L$ vs. $V$, and improves on it by determining how $H$ is incorporated and by accounting for the granularity through the distribution of the grain diameter $D$. %Although a dependence on $H$ is clear \cite{johnson2017drop}, sometimes the ratio $H/L$ can perform even worse than just $L$ by itself \cite{staron2009understanding}, suggesting more work is needed to understand the correct dependence of $L$ on $H$. On the other hand, 
The extreme scale separation between the grain size (typically ranging from $\sim 10^{-3}$ m to $\sim 10^{0}$ m, see Supplementary Material $SM$ \cite{seeSM}) and either $H$ ($\sim 10^2$ m) or $L$ ($\sim 10^3$ m) suggests it is a negligible parameter, an implicit assumption in many continuum approaches. And yet it is an integral parameter in prominent granular rheologies \cite{jop2006constitutive,forterre2008flows}, in understanding laboratory granular experiments \cite{lajeunesse2005granular}, and in improving rheological models to replicate landslides \cite{pirulli2008results}. Grain size has been used implicitly in rheologies that predict basal (bottom) friction \cite{parez2015long}, and explicitly in studies of air drag \cite{kesseler2020grain}, but its influence on runout has only rarely been studied systematically \cite{fei2020particle}, often being ignored altogether \cite{johnson2017drop,lucas2014frictional} or even declared irrelevant \cite{davies1999runout}.

Here, by explicitly and systematically accounting for $D$ and $H$, we not only improve the correlation between $L$, $H$, and landslide size, but also weaken the boundaries between the heretofore distinct flows. The implications of this result are twofold. First, the underlying physical mechanisms that control laboratory granular landslides and a number of natural landslides are apparently the same, potentially eliminating the need for emergent mechanisms to explain the runout of some large-scale landslides \cite{johnson2016reduction,davies1999runout}. Second, laboratory-scale experiments can thus be fruitfully used to systematically investigate the quantitative behavior of even large-scale landslides, despite a widespread view to the contrary \cite{iverson2015scaling,delannay2017granular}. Our simple experiments and analysis thus serve to both unite disparate fields and potentially enlarge others \cite{companionPRE}.

\begin{figure}[t!]
\centering
\includegraphics[width=1.0\linewidth]{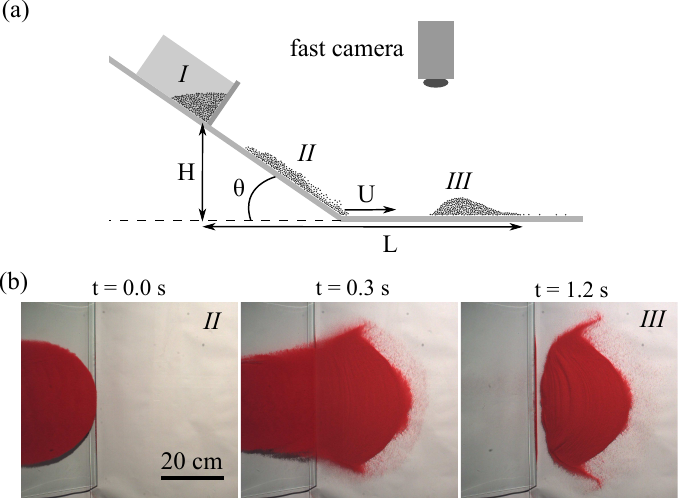}
\caption{(a) Side-view schematic of the experimental setup with (b) overhead images at intermediate ($II$) and final ($III$) times of the motion of $M \simeq$ 1300 g of artificially coloured red sand ($\langle D \rangle \simeq 250$ $\mu$m). The grains are released from a rectangular box ($I$) of width 15 cm via a sliding sluice gate at the front and slide down a flat glass plate (80 cm long and 65 cm wide), inclined at an angle $\theta \simeq 34^\circ$, and eventually reach the junction ($II$) before coming to rest on a level flat glass plate (125 cm by 125 cm). The grain motion is observed with an overhead fast camera to determine the front speed $U$ just after the junction. We use the frontmost position of the main mass to determine the runout $L$ in our experiments.}
\vspace{-0em}
\label{setup}
\end{figure}

We begin with laboratory experiments for which we systematically vary the fall height, the landslide size, and the grain size. The experimental setup is a simplified version of a natural landslide consisting of a slope, a flat section, grains, and a container to house the grains before releasing them by rapidly raising a sliding metal gate. See Fig.~\ref{setup}a for a schematic of the experimental setup. Natural landslides vary widely in the details of their initial conditions, and we make no attempt to replicate the specific features of any particular landslide geometry. While these details can affect some aspects of landslide dynamics, the runout distance itself is relatively insensitive \cite{lucas2011influence}. %Following the traditional procedure in landslide research, we define an effective friction as $\mu = H/L$, where $H$ is the vertical fall height and $L$ is the horizontal travel distance \cite{legros2002mobility,lucas2014frictional,delannay2017granular}. In contrast to previous studies, however, we define both $H$ and $L$ with reference to the front position for the beginning and end of the landslide event.
In contrast to some previous studies \cite{lucas2014frictional}, we define both $H$ and $L$ with reference to the front position for both the beginning and end of the landslide event.

\begin{figure*}
\centering
\includegraphics[width=1\linewidth]{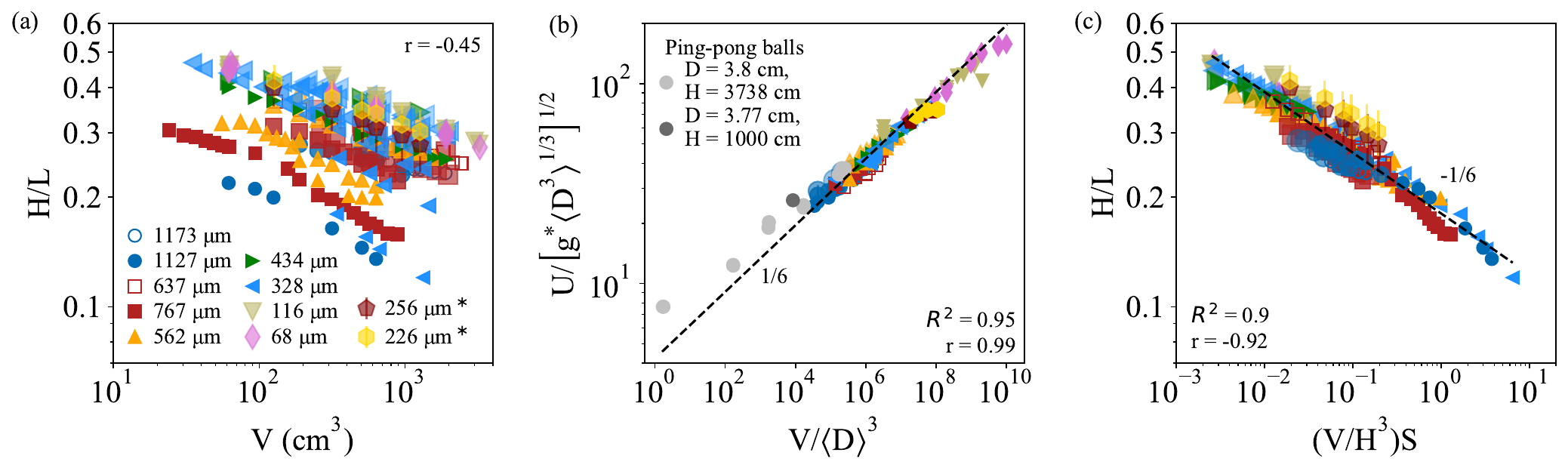}
\vspace{-2em}
\caption{Plots of runout and $U$ from the laboratory experiments with corresponding Spearman correlation coefficients $r$ for all data in each subplot. (a) A traditional plot of $(H/L)$ vs. $V$ for laboratory experiments with different $\langle D \rangle$ and $H$ ($5$ cm $\lesssim H \lesssim 37$ cm, data symbols for larger $H$ are larger and more transparent). There is an overall negative trend in keeping with natural landslides \cite{legros2002mobility}, but the data show scatter and a clear dependence on $D$ and $H$. Grain sizes in the legend refer to $\langle D \rangle$. The two grains marked by $^\ast$ are bidisperse mixtures with large $S \simeq 4$. (b) A plot of the front speeds $U$ vs. $V$ from experiments and ping-pong ball experiments \cite{mcelwaine2001ping,nishimura1998ping,kosugi1995table} normalized using the grain-size distribution ($\langle D^3\rangle$, $\langle D \rangle$). All dashed lines ($--$) are a power-law fit with exponent 1/6. The best fit exponent is 0.143$\pm$0.002. (c) The runout data collapse by taking into account both $D$ and $H$ in the normalization and go as $H/L \sim [(V/H^3) S]^{-1/6}$ ($--$). The best fit exponent is -0.156$\pm$0.003.}
\label{laboratoryCollapse}
\end{figure*}

For each experiment we measure the total mass $M$ with a simple scale and gently sprinkle the grains into the rectangular box so that the surface is level. We then measure the total volume $V$ of the initial grain pile using a standard measuring tape, which we also use to measure $H$ and $L$. Because individual grains can escape and travel farther than the bulk, we identify the final front position as the frontmost position where a layer of grains is still in contact with the main mass (see Fig.~\ref{setup}b). In addition, we also measure the landslide front position and speed $U$ using an overhead fast camera (Phantom v641) at a frame rate of 100 Hz. To determine $U$ we used a standard image processing tool (ImageJ) to manually track the front position of the landslide, which is easily distinguished in the experiments (see Fig.~\ref{setup}b and $SM$ \cite{seeSM}). Our grains are relatively spherical glass beads which have been roughly pre-sorted by the manufacturer according to diameter $D$, ranging from $\sim$ 45 $\mu$m to $\sim$ 1.5 mm, and we characterize the size and shape distributions of the grains using an imaging technique \cite{cerbus2021landslide} (see also $SM$ \cite{seeSM}). We use the mass-weighted distribution $p(D)$ to determine average quantities such as $\langle D \rangle$ and $\langle D^3 \rangle$. %This yields the frequency distribution $p_f(D)$, which we then convert into the mass-weighted distribution $p(D) \propto p_f(D) \times (4 \pi \rho/3) (D/2)^3$, where $\rho$ is the grain material mass density. We calculate average quantities such as $\langle D \rangle$ and $\langle D^3 \rangle$ using $p(D)$. %For distributions that are nearly symmetric, averages calculated using $p(D)$ and $p_f(D)$ are approximately the same $\langle D \rangle_f \simeq \langle D \rangle$, but we also include data with a bimodal size distribution, for which the frequency-weighted and mass-weighted averages are not equivalent (see SM \cite{seeSM}). We found empirically that using $p(D)$ yields better results for such distributions. 
% Electrostatic effects can be prominent for very small particle sizes ($\lesssim 90$ $\mu$m) \cite{shinbrot2012electrostatic}. In the present experiments all average grain sizes but one were larger than this.
We observed no substantial influence of electrostatic effects on the landslide runout distance.

% \section{Landslides in the dense limit}
% \label{sec:denseLimit}

% \subsection{Re-analysis including $\langle D \rangle$}
% \label{sec:analysisDense}

In Fig.~\ref{laboratoryCollapse}a we show different experimental curves of Heim's ratio $H/L$ vs. $V$ (we consider even lower $V$ in the accompanying article \cite{companionPRE}). In our experiments we systematically vary $\langle D \rangle$ and $H$, and perform experiments for a large range of $V$. This creates a jumble of data. Varying $H$ or $\langle D \rangle$ creates a new curve, but it is similar in shape, follows the classical trend, and is only shifted vertically. Increasing $H$ creates a new curve above, and increasing $\langle D \rangle$ creates a new curve below. Such influence is not surprising and has been previously noted \cite{johnson2017drop,fei2020particle}, but to our knowledge the quantitative dependence has never been elucidated. Inspired by the apparent similarity of the individual curves and a systematic dependence on $H$ and $\langle D \rangle$, we seek a self-similar function for the dependence of $L$ on $V$, $H$, and $D$. 

We find inspiration for this solution from the results of large-scale experiments of up to 550,000 ping-pong balls released on a ski jump and in the laboratory \cite{mcelwaine2001ping,nishimura1998ping,kosugi1995table}. These experiments found that the front speed $U$ scales with the size of the system, represented by the number of ping-pong balls $N$, as $U \sim N^{1/6} \sim V^{1/6}$. Here we extend their analysis by determining the dependence on $D$ and $H$. %First using $V$ to represent the system size, we plot $U$ vs. $V$ for the ping-pong ball data and our own data and confirm $U \propto V^{1/6}$ (see inset to Fig.~\ref{laboratoryCollapse}b). However, there is still significant scatter in the experimental data and a gap between our experiments and theirs. 
We found that while our laboratory experiments, like the ping-pong ball experiments, roughly yield $U \sim V^{1/6}$ (see $SM$ \cite{seeSM}), for proper comparison we must non-dimensionalize $U$ and $V$. This requires at least one length scale $l$ (the velocity can be scaled with $\sqrt{g l}$, where $g$ is the gravitational acceleration. We determined that $U$ does not depend on $H$ %, since their experiments spanned 10 m $\leq H \leq $ 50 m, and yet their data follow the same curve. Likewise it was not possible to collapse our data using $H$, which in our (front speed) experiments varied from 16 cm $\lesssim H \lesssim$ 36 cm (see $SM$ \cite{seeSM}). 
(see $SM$ \cite{seeSM}), so we instead turn to $D$. Because the mobility of granular flow has been linked to its particle size distribution ($p(D)$) in laboratory experiments \cite{moro2010large}, two-dimensional simulations \cite{linares2007increased}, and in granular slumping experiments and simulations \cite{lai2017collapse}, we anticipate that $U$ may depend on various moments of $p(D)$. We choose the first ($\langle D \rangle$) and third ($\langle D^3 \rangle$) moments, since the latter is related to the asymmetry of $p(D)$. In most of the laboratory and ping-pong experiments their ratio is $S \equiv \langle D^3 \rangle / \langle D \rangle^3 \sim 1$, so we performed several experiments with mixed grains to obtain more asymmetric distributions ($S \sim 4$) and found substantially better collapse when we normalize $U$ using $\langle D^3 \rangle$ and $V$ using $\langle D \rangle$ (see $SM$ \cite{seeSM}). We thus define a characteristic speed $\sqrt{g^* \langle D^3 \rangle ^{1/3}}$, where $g^* = g(1-\rho_{\rm{air}}/\rho)(\sin \theta - \mu_{\rm{sliding}} \cos \theta)$, $g$ is the gravitational acceleration, $\rho_{\rm{air}}$ is the air density, and $\mu_{\rm{sliding}}$ is the sliding friction coefficient from the literature \cite{pongo2021flow,inaba2017effect}. We plot the result in Fig.~\ref{laboratoryCollapse}b. Not only is the laboratory collapse improved (a fit to a 1/6 power-law changes from $R^2 \simeq$ 0.85 to $R^2 \simeq$ 0.95), but the laboratory and ping-pong data are brought into close correspondence. %While accounting for the granularity was necessary to produce the collapse in Fig.~\ref{laboratoryCollapse}b, because their grain size distributions are not heavily skewed ($\langle D^3 \rangle^{1/3}/\langle D \rangle \simeq 1$), %\footnote{We note that the typical definition of skewness, $\mu_3$, is not equivalent to but is related to the ratio $\langle D^3 \rangle^{1/3}/\langle D \rangle$.},
%it is difficult to determine which combination is appropriate. We confirm our choice here when we later consider the field data. 

We connect this result for the motion of the landslide with its runout using a simple argument. We estimate $L$ using the front speed $U$ so that $L \sim UT$, where $T$ is a characteristic time we assume to go as $T \sim \sqrt{H/g^*}$ as observed in granular slumping experiments \cite{lajeunesse2005granular}. We thus estimate Heim's ratio as $H/L = H/UT$, yielding
\begin{equation}
    \frac{H}{L} \sim \frac{H \sqrt{g^* \langle D \rangle}}{\sqrt{H g^* \langle D^3 \rangle^{1/3} V^{1/3}}} \sim \left[ \left( \frac{V}{H^3} \right) S \right]^{-1/6},
    \label{eq:similarity}
\end{equation}
where only the final term $S = \langle D^3 \rangle/\langle D \rangle^3$ depends on the granularity through the asymmetry of $p(D)$. We test this prediction in Fig.~\ref{laboratoryCollapse}c, in which we observe the remarkable collapse engendered by the scaling of Eq.~(\ref{eq:similarity}), including even the exponent $-1/6$. %While accounting for the granularity was essential to yield the collapse in Fig.~\ref{laboratoryCollapse}b, it is $H$ which does the work to yield the collapse in Fig.~\ref{laboratoryCollapse}c.
Our simple argument together with Fig.~\ref{laboratoryCollapse} not only demonstrate for the first time that the quantitative dependence of the runout distance on $H$ is through the square-root ($\propto \sqrt{H}$) but that the runout speed and distance also depend on the grain distribution asymmetry through $S$. We found that the collapse in Figs.~\ref{laboratoryCollapse}b,c does not hold for small values of $V$, with a lower limit that depends on $H$ and $D$ \cite{companionPRE}.

\begin{figure*}[t!]
\centering
\includegraphics[width=1\linewidth]{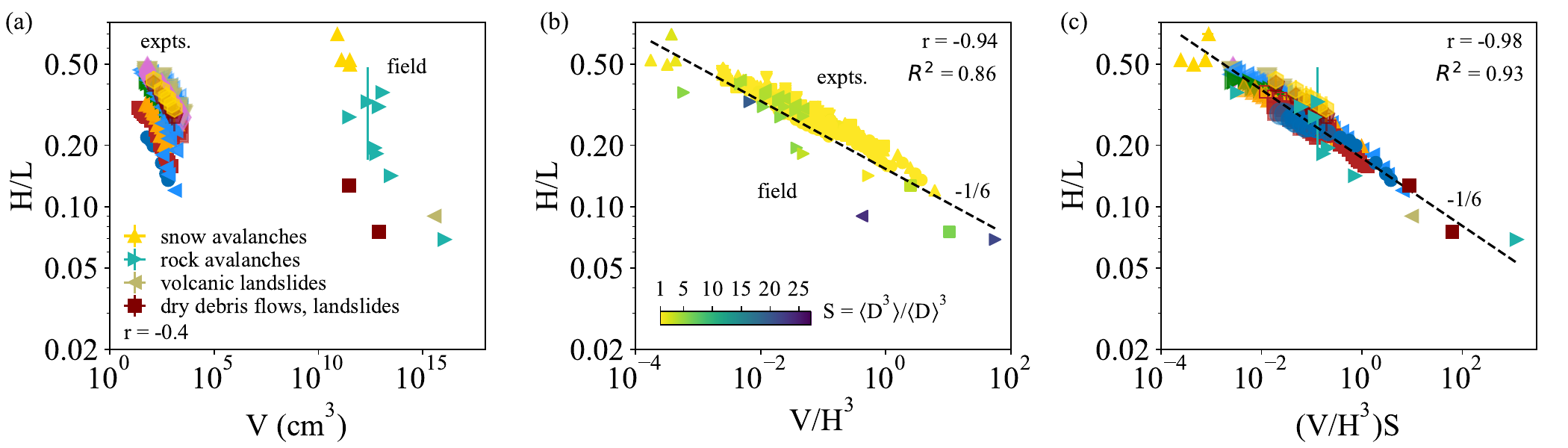}
\vspace{-2em}
\caption{Plots of runout from the laboratory, dry debris flows and landslides \cite{legros2002mobility}, rock avalanches \cite{heim1932bergsturz}, snow avalanches \cite{atwater1954snow}, volcanic landslides \cite{newhall2000} with corresponding Spearman correlation coefficients $r$ for all data in each subplot. We sampled laboratory data points evenly (logarithmically) spaced and with a number equal to the field data for the correlation and goodness of fit ($R^2$). (a) A traditional plot of $H/L$ vs. $V$ shows large scatter and separates the field and laboratory data. (b) Re-plotting $H/L$ vs. $V/H^3$ yields a substantial improvement and reveals the dependence on $\propto H^{1/2}$. The dashed line ($--$) is the predicted -1/6 power-law. The best fit exponent is -0.164$\pm$0.012. The values of $S$ are indicated by the marker color with larger values tending to be below. (c) Re-plotting with the new scaling which includes the granularity yields excellent accord with a single, nearly universal curve $\sim [(V/H^3) S]^{-1/6}$ ($--$). The best fit exponent is -0.158$\pm$0.008. Error bars are determined from the spread of values for $V$, $H$, $L$, and $D$ (see $SM$ \cite{seeSM}).
}
\label{literatureCollapse}
\vspace{0em}
\end{figure*}

Now we apply our scaling to naturally occurring landslides, the ultimate motivation for our experiments. Unlike our relatively simple laboratory experiments, natural landslides are influenced by a host of additional parameters \cite{aaron2019rock,iverson2015scaling}, some of which we cannot account for with a purely granular flow, so that we might not \emph{a priori} expect our simple scaling to be relevant. Indeed Fig.~\ref{literatureCollapse}a, a traditional plot of $\mu = H/L$ vs. $V$, visually demonstrates the apparent irrelevance of laboratory experiments to natural landslides. Even focusing on the natural landslides only, which consist of dry debris flows \cite{legros2002mobility}, rock avalanches \cite{heim1932bergsturz,mcsaveney2002recent,pfiffner2022flims}, snow avalanches \cite{atwater1954snow}, and volcanic landslides \cite{newhall2000}, we can observe the large scatter which has historically encouraged separate treatment for each type of mass movement. In order to test our scaling prediction (Eq.~(\ref{eq:similarity})), we use previously measured grain size distributions from literature sources, which are often extremely asymmetric, to determine $S$ (see $SM$ \cite{seeSM}).

First we consider the possible improvement from correctly accounting for $H$, temporarily ignoring the granularity. As Fig.~\ref{literatureCollapse}b shows, when the mess of apparently distinct data (Fig.~\ref{literatureCollapse}a) are plotted vs. $V/H^3$, we observe a striking improvement in the collapse of the data, including the apparent affinity between small-scale laboratory experiments and field data which is hidden by the traditional plot (Fig.~\ref{literatureCollapse}a). Likewise the data broadly conform to a power-law with exponent -1/6 as predicted by our simple argument (Eq.~(\ref{eq:similarity})). We thus confirm that the dependence on $H$ through its square-root is a ubiquitous feature common to landslides over a large range of drop heights, $10^{0}$ cm $\lesssim H \lesssim 10^5$ cm.

Next we turn to the landslide granularity. In Fig.~\ref{literatureCollapse}b we have plotted the data with a color corresponding to its value of the $S$. While for most of the laboratory data $S \sim 1$, for the field data it can reach values of $S \sim $ 27. Moreover the data with large values of this ratio tend to fall below while those with smaller values are above. The skewness of the grain size distribution apparently increases landslide mobility, a result anticipated by two-dimensional simulations \cite{linares2007increased} and laboratory experiments \cite{moro2010large} with grain distributions having a large bidispersity. When we include this ratio in the normalization of the system size as in Fig.~\ref{laboratoryCollapse}c and according to Eq.~(\ref{eq:similarity}), we find a significant improvement to the collapse of all the data, which is noticeably better than representations that exclude $D$ (see $SM$ \cite{seeSM}). In particular, including the granularity (skewness) brings the field data into even closer correspondence with the laboratory experiments.

The success of Fig.~\ref{literatureCollapse}c does not signify that other factors, such as moisture \cite{iverson2000acute} or topography \cite{aaron2019rock}, do not influence landslides. Indeed, the scatter of the data may in fact be manifestations of these other parameters. However, Fig.~\ref{literatureCollapse}c suggests that it is $V$, $H$, and $D$ which primarily determine the behavior of $L$. %It is particularly noteworthy that our smooth substrate experiments are in good accord with field data, despite their varied topography \cite{lucas2011influence}, although this result is already anticipated by the good agreement between natural landslides and simulations on smooth substrates \cite{campbell1995large,johnson2016reduction}.
Previous work has often tried to detect the influence of other parameters using the traditional plot of $H/L$ vs $V$ \cite{legros2002mobility,johnson2016reduction,johnson2017drop}, but our improved scaling now provides a framework within which to determine the role of other parameters systematically using laboratory experiments. %Since we have demonstrated that the runout of large-scale mass movements of all kinds behaves quantitatively in the same way as each other and as our laboratory experiments, the role of other relevant parameters can be investigated using systematic experiments and scaling analysis as done here, and the need for special emergent mechanisms largely disappears.
On a practical level, the scaling result in Fig.~\ref{literatureCollapse}c also provides an improved method for estimating landslide runout hazard \cite{keefer2007assessing}. Building on traditional methods which use historical data to predict $V$ and $H$ \cite{cardinali2002geomorphological}, with additional information about a region's associated grain size distribution, an estimate can be made of $S$ to predict $L$.

In summary, we have performed an extensive experimental study of laboratory landslide runout combined with a scaling prediction that has led to a quantitative understanding of the dominant parameters that control even natural landslides. We found that the system size $V$, the fall height $H$, and the granularity $D$, while certainly not the only parameters, are among the most important parameters that determine the behavior of granular laboratory experiments, rock avalanches, snow avalanches, landslides, and dry debris flows. In demonstrating this we have found an impressive link between seemingly disparate systems: their granular nature. %With our findings here as a starting point, future work can begin to add other relevant parameters such as moisture \cite{iverson2000acute} and terrain roughness \cite{iverson2015scaling}, and perform similar systematic experiments and scaling analysis to elucidate the full dependence of landslide runout on its rich set of control variables \cite{companionPRE}. Further improvements to the collapse of the normalized runout data, as found here, will enhance important quantitative tools for estimating landslide hazard \cite{keefer2007assessing}. 
%Stanisław Jerzy Lec wrote, regarding collective guilt, that ``No snowflake in an avalanche ever feels responsible" to suggest that the individual often feels independent from the collective \cite{lec1962unkempt}. Questions about guilt aside, we have found it to be at least inversely true that the collective behavior of landslides, specifically the runout, is not independent of the properties of its individual grains.

\emph{Acknowledgments.} R.T.C. gratefully acknowledges funding from the Horizon 2020 program under the Marie Skłodowska-Curie Action Individual Fellowship (MSCAIF) No. 793507, and H.K. acknowledges support from the Institut Universitaire de France. We thank Alexis Petit and Alioune Mbodji for their help in performing some of the experiments. %{\bf Author contributions:} R.T.C., T.F., and H.K. conceived the study; R.T.C. conducted the experiments and analyzed the laboratory and field data with input from T.F. and H.K.; R.T.C., T.F., and H.K. discussed the results; R.T.C. wrote the manuscript with input from T.F. and H.K. { \bf Competing interests:} The authors declare no competing interests. 

\bibliography{landslideFrictionCollapsePRL}

\begin{thebibliography}{42}%
\makeatletter
\providecommand \@ifxundefined [1]{%
 \@ifx{#1\undefined}
}%
\providecommand \@ifnum [1]{%
 \ifnum #1\expandafter \@firstoftwo
 \else \expandafter \@secondoftwo
 \fi
}%
\providecommand \@ifx [1]{%
 \ifx #1\expandafter \@firstoftwo
 \else \expandafter \@secondoftwo
 \fi
}%
\providecommand \natexlab [1]{#1}%
\providecommand \enquote  [1]{``#1''}%
\providecommand \bibnamefont  [1]{#1}%
\providecommand \bibfnamefont [1]{#1}%
\providecommand \citenamefont [1]{#1}%
\providecommand \href@noop [0]{\@secondoftwo}%
\providecommand \href [0]{\begingroup \@sanitize@url \@href}%
\providecommand \@href[1]{\@@startlink{#1}\@@href}%
\providecommand \@@href[1]{\endgroup#1\@@endlink}%
\providecommand \@sanitize@url [0]{\catcode `\\12\catcode `\$12\catcode
  `\&12\catcode `\#12\catcode `\^12\catcode `\_12\catcode `\%12\relax}%
\providecommand \@@startlink[1]{}%
\providecommand \@@endlink[0]{}%
\providecommand \url  [0]{\begingroup\@sanitize@url \@url }%
\providecommand \@url [1]{\endgroup\@href {#1}{\urlprefix }}%
\providecommand \urlprefix  [0]{URL }%
\providecommand \Eprint [0]{\href }%
\providecommand \doibase [0]{https://doi.org/}%
\providecommand \selectlanguage [0]{\@gobble}%
\providecommand \bibinfo  [0]{\@secondoftwo}%
\providecommand \bibfield  [0]{\@secondoftwo}%
\providecommand \translation [1]{[#1]}%
\providecommand \BibitemOpen [0]{}%
\providecommand \bibitemStop [0]{}%
\providecommand \bibitemNoStop [0]{.\EOS\space}%
\providecommand \EOS [0]{\spacefactor3000\relax}%
\providecommand \BibitemShut  [1]{\csname bibitem#1\endcsname}%
\let\auto@bib@innerbib\@empty
%</preamble>
\bibitem [{\citenamefont {Keefer}\ and\ \citenamefont
  {Larsen}(2007)}]{keefer2007assessing}%
  \BibitemOpen
  \bibfield  {author} {\bibinfo {author} {\bibfnamefont {D.~K.}\ \bibnamefont
  {Keefer}}\ and\ \bibinfo {author} {\bibfnamefont {M.~C.}\ \bibnamefont
  {Larsen}},\ }\bibfield  {title} {\bibinfo {title} {Assessing landslide
  hazards},\ }\href@noop {} {\bibfield  {journal} {\bibinfo  {journal}
  {Science}\ }\textbf {\bibinfo {volume} {316}},\ \bibinfo {pages} {1136}
  (\bibinfo {year} {2007})}\BibitemShut {NoStop}%
\bibitem [{\citenamefont {Perkins}(2012)}]{perkins2012death}%
  \BibitemOpen
  \bibfield  {author} {\bibinfo {author} {\bibfnamefont {S.}~\bibnamefont
  {Perkins}},\ }\bibfield  {title} {\bibinfo {title} {Death toll from
  landslides vastly underestimated},\ }\href@noop {} {\bibfield  {journal}
  {\bibinfo  {journal} {Nature News}\ } (\bibinfo {year} {2012})}\BibitemShut
  {NoStop}%
\bibitem [{\citenamefont {Newhall}(2000)}]{newhall2000}%
  \BibitemOpen
  \bibfield  {author} {\bibinfo {author} {\bibfnamefont {C.~G.}\ \bibnamefont
  {Newhall}},\ }\bibfield  {title} {\bibinfo {title} {Mount {St}. {Helens},
  {Master} {Teacher}},\ }\href {https://doi.org/10.1126/science.288.5469.1181}
  {\bibfield  {journal} {\bibinfo  {journal} {Science}\ }\textbf {\bibinfo
  {volume} {288}},\ \bibinfo {pages} {1181} (\bibinfo {year}
  {2000})}\BibitemShut {NoStop}%
\bibitem [{\citenamefont {Legros}(2002)}]{legros2002mobility}%
  \BibitemOpen
  \bibfield  {author} {\bibinfo {author} {\bibfnamefont {F.}~\bibnamefont
  {Legros}},\ }\bibfield  {title} {\bibinfo {title} {The mobility of
  long-runout landslides},\ }\href@noop {} {\bibfield  {journal} {\bibinfo
  {journal} {Eng. {G}eol.}\ }\textbf {\bibinfo {volume} {63}},\ \bibinfo
  {pages} {301} (\bibinfo {year} {2002})}\BibitemShut {NoStop}%
\bibitem [{\citenamefont {Hungr}\ \emph {et~al.}(2014)\citenamefont {Hungr},
  \citenamefont {Leroueil},\ and\ \citenamefont {Picarelli}}]{hungr2014varnes}%
  \BibitemOpen
  \bibfield  {author} {\bibinfo {author} {\bibfnamefont {O.}~\bibnamefont
  {Hungr}}, \bibinfo {author} {\bibfnamefont {S.}~\bibnamefont {Leroueil}},\
  and\ \bibinfo {author} {\bibfnamefont {L.}~\bibnamefont {Picarelli}},\
  }\bibfield  {title} {\bibinfo {title} {The varnes classification of landslide
  types, an update},\ }\href@noop {} {\bibfield  {journal} {\bibinfo  {journal}
  {Landslides}\ }\textbf {\bibinfo {volume} {11}},\ \bibinfo {pages} {167}
  (\bibinfo {year} {2014})}\BibitemShut {NoStop}%
\bibitem [{\citenamefont {Atwater}(1954)}]{atwater1954snow}%
  \BibitemOpen
  \bibfield  {author} {\bibinfo {author} {\bibfnamefont {M.~M.}\ \bibnamefont
  {Atwater}},\ }\bibfield  {title} {\bibinfo {title} {Snow avalanches},\
  }\href@noop {} {\bibfield  {journal} {\bibinfo  {journal} {Sci. Am.}\
  }\textbf {\bibinfo {volume} {190}},\ \bibinfo {pages} {26} (\bibinfo {year}
  {1954})}\BibitemShut {NoStop}%
\bibitem [{\citenamefont {Magnarini}\ \emph {et~al.}(2019)\citenamefont
  {Magnarini}, \citenamefont {Mitchell}, \citenamefont {Grindrod},
  \citenamefont {Goren},\ and\ \citenamefont
  {Schmitt}}]{magnarini2019longitudinal}%
  \BibitemOpen
  \bibfield  {author} {\bibinfo {author} {\bibfnamefont {G.}~\bibnamefont
  {Magnarini}}, \bibinfo {author} {\bibfnamefont {T.~M.}\ \bibnamefont
  {Mitchell}}, \bibinfo {author} {\bibfnamefont {P.~M.}\ \bibnamefont
  {Grindrod}}, \bibinfo {author} {\bibfnamefont {L.}~\bibnamefont {Goren}},\
  and\ \bibinfo {author} {\bibfnamefont {H.~H.}\ \bibnamefont {Schmitt}},\
  }\bibfield  {title} {\bibinfo {title} {Longitudinal ridges imparted by
  high-speed granular flow mechanisms in martian landslides},\ }\href@noop {}
  {\bibfield  {journal} {\bibinfo  {journal} {Nat. Commun.}\ }\textbf {\bibinfo
  {volume} {10}},\ \bibinfo {pages} {1} (\bibinfo {year} {2019})}\BibitemShut
  {NoStop}%
\bibitem [{\citenamefont {Iverson}\ and\ \citenamefont
  {LaHusen}(1989)}]{iverson1989dynamic}%
  \BibitemOpen
  \bibfield  {author} {\bibinfo {author} {\bibfnamefont {R.~M.}\ \bibnamefont
  {Iverson}}\ and\ \bibinfo {author} {\bibfnamefont {R.~G.}\ \bibnamefont
  {LaHusen}},\ }\bibfield  {title} {\bibinfo {title} {Dynamic pore-pressure
  fluctuations in rapidly shearing granular materials},\ }\href@noop {}
  {\bibfield  {journal} {\bibinfo  {journal} {Science}\ }\textbf {\bibinfo
  {volume} {246}},\ \bibinfo {pages} {796} (\bibinfo {year}
  {1989})}\BibitemShut {NoStop}%
\bibitem [{\citenamefont {Iverson}\ \emph {et~al.}(2000)\citenamefont
  {Iverson}, \citenamefont {Reid}, \citenamefont {Iverson}, \citenamefont
  {LaHusen}, \citenamefont {Logan}, \citenamefont {Mann},\ and\ \citenamefont
  {Brien}}]{iverson2000acute}%
  \BibitemOpen
  \bibfield  {author} {\bibinfo {author} {\bibfnamefont {R.~M.}\ \bibnamefont
  {Iverson}}, \bibinfo {author} {\bibfnamefont {M.}~\bibnamefont {Reid}},
  \bibinfo {author} {\bibfnamefont {N.~R.}\ \bibnamefont {Iverson}}, \bibinfo
  {author} {\bibfnamefont {R.}~\bibnamefont {LaHusen}}, \bibinfo {author}
  {\bibfnamefont {M.}~\bibnamefont {Logan}}, \bibinfo {author} {\bibfnamefont
  {J.}~\bibnamefont {Mann}},\ and\ \bibinfo {author} {\bibfnamefont
  {D.}~\bibnamefont {Brien}},\ }\bibfield  {title} {\bibinfo {title} {Acute
  sensitivity of landslide rates to initial soil porosity},\ }\href@noop {}
  {\bibfield  {journal} {\bibinfo  {journal} {Science}\ }\textbf {\bibinfo
  {volume} {290}},\ \bibinfo {pages} {513} (\bibinfo {year}
  {2000})}\BibitemShut {NoStop}%
\bibitem [{\citenamefont {Faug}\ \emph {et~al.}(2008)\citenamefont {Faug},
  \citenamefont {Gauer}, \citenamefont {Lied},\ and\ \citenamefont
  {Naaim}}]{faug2008overrun}%
  \BibitemOpen
  \bibfield  {author} {\bibinfo {author} {\bibfnamefont {T.}~\bibnamefont
  {Faug}}, \bibinfo {author} {\bibfnamefont {P.}~\bibnamefont {Gauer}},
  \bibinfo {author} {\bibfnamefont {K.}~\bibnamefont {Lied}},\ and\ \bibinfo
  {author} {\bibfnamefont {M.}~\bibnamefont {Naaim}},\ }\bibfield  {title}
  {\bibinfo {title} {Overrun length of avalanches overtopping catching dams:
  Cross-comparison of small-scale laboratory experiments and observations from
  full-scale avalanches},\ }\href@noop {} {\bibfield  {journal} {\bibinfo
  {journal} {J. Geophys. Res. Earth Surface}\ }\textbf {\bibinfo {volume}
  {113}} (\bibinfo {year} {2008})}\BibitemShut {NoStop}%
\bibitem [{\citenamefont {Miller}\ \emph {et~al.}(2017)\citenamefont {Miller},
  \citenamefont {Andy~Take}, \citenamefont {Mulligan},\ and\ \citenamefont
  {McDougall}}]{miller2017tsunamis}%
  \BibitemOpen
  \bibfield  {author} {\bibinfo {author} {\bibfnamefont {G.~S.}\ \bibnamefont
  {Miller}}, \bibinfo {author} {\bibfnamefont {W.}~\bibnamefont {Andy~Take}},
  \bibinfo {author} {\bibfnamefont {R.~P.}\ \bibnamefont {Mulligan}},\ and\
  \bibinfo {author} {\bibfnamefont {S.}~\bibnamefont {McDougall}},\ }\bibfield
  {title} {\bibinfo {title} {Tsunamis generated by long and thin granular
  landslides in a large flume},\ }\href@noop {} {\bibfield  {journal} {\bibinfo
   {journal} {J. Geophys. Res. Oceans}\ }\textbf {\bibinfo {volume} {122}},\
  \bibinfo {pages} {653} (\bibinfo {year} {2017})}\BibitemShut {NoStop}%
\bibitem [{\citenamefont {Daerr}\ and\ \citenamefont
  {Douady}(1999)}]{daerr1999two}%
  \BibitemOpen
  \bibfield  {author} {\bibinfo {author} {\bibfnamefont {A.}~\bibnamefont
  {Daerr}}\ and\ \bibinfo {author} {\bibfnamefont {S.}~\bibnamefont {Douady}},\
  }\bibfield  {title} {\bibinfo {title} {Two types of avalanche behaviour in
  granular media},\ }\href@noop {} {\bibfield  {journal} {\bibinfo  {journal}
  {Nature}\ }\textbf {\bibinfo {volume} {399}},\ \bibinfo {pages} {241}
  (\bibinfo {year} {1999})}\BibitemShut {NoStop}%
\bibitem [{\citenamefont {Iverson}(2015)}]{iverson2015scaling}%
  \BibitemOpen
  \bibfield  {author} {\bibinfo {author} {\bibfnamefont {R.~M.}\ \bibnamefont
  {Iverson}},\ }\bibfield  {title} {\bibinfo {title} {Scaling and design of
  landslide and debris-flow experiments},\ }\href@noop {} {\bibfield  {journal}
  {\bibinfo  {journal} {Geomorphology}\ }\textbf {\bibinfo {volume} {244}},\
  \bibinfo {pages} {9} (\bibinfo {year} {2015})}\BibitemShut {NoStop}%
\bibitem [{\citenamefont {Davies}\ and\ \citenamefont
  {McSaveney}(1999)}]{davies1999runout}%
  \BibitemOpen
  \bibfield  {author} {\bibinfo {author} {\bibfnamefont {T.}~\bibnamefont
  {Davies}}\ and\ \bibinfo {author} {\bibfnamefont {M.}~\bibnamefont
  {McSaveney}},\ }\bibfield  {title} {\bibinfo {title} {Runout of dry granular
  avalanches},\ }\href@noop {} {\bibfield  {journal} {\bibinfo  {journal} {Can.
  Geotech. J.}\ }\textbf {\bibinfo {volume} {36}},\ \bibinfo {pages} {313}
  (\bibinfo {year} {1999})}\BibitemShut {NoStop}%
\bibitem [{\citenamefont {Delannay}\ \emph {et~al.}(2017)\citenamefont
  {Delannay}, \citenamefont {Valance}, \citenamefont {Mangeney}, \citenamefont
  {Roche},\ and\ \citenamefont {Richard}}]{delannay2017granular}%
  \BibitemOpen
  \bibfield  {author} {\bibinfo {author} {\bibfnamefont {R.}~\bibnamefont
  {Delannay}}, \bibinfo {author} {\bibfnamefont {A.}~\bibnamefont {Valance}},
  \bibinfo {author} {\bibfnamefont {A.}~\bibnamefont {Mangeney}}, \bibinfo
  {author} {\bibfnamefont {O.}~\bibnamefont {Roche}},\ and\ \bibinfo {author}
  {\bibfnamefont {P.}~\bibnamefont {Richard}},\ }\bibfield  {title} {\bibinfo
  {title} {Granular and particle-laden flows: from laboratory experiments to
  field observations},\ }\href@noop {} {\bibfield  {journal} {\bibinfo
  {journal} {Journal of Physics D: Applied Physics}\ }\textbf {\bibinfo
  {volume} {50}},\ \bibinfo {pages} {053001} (\bibinfo {year}
  {2017})}\BibitemShut {NoStop}%
\bibitem [{\citenamefont {Lucas}\ \emph {et~al.}(2014)\citenamefont {Lucas},
  \citenamefont {Mangeney},\ and\ \citenamefont
  {Ampuero}}]{lucas2014frictional}%
  \BibitemOpen
  \bibfield  {author} {\bibinfo {author} {\bibfnamefont {A.}~\bibnamefont
  {Lucas}}, \bibinfo {author} {\bibfnamefont {A.}~\bibnamefont {Mangeney}},\
  and\ \bibinfo {author} {\bibfnamefont {J.~P.}\ \bibnamefont {Ampuero}},\
  }\bibfield  {title} {\bibinfo {title} {Frictional velocity-weakening in
  landslides on earth and on other planetary bodies},\ }\href@noop {}
  {\bibfield  {journal} {\bibinfo  {journal} {Nat. Commun.}\ }\textbf {\bibinfo
  {volume} {5}},\ \bibinfo {pages} {1} (\bibinfo {year} {2014})}\BibitemShut
  {NoStop}%
\bibitem [{\citenamefont {Johnson}\ and\ \citenamefont
  {Campbell}(2017)}]{johnson2017drop}%
  \BibitemOpen
  \bibfield  {author} {\bibinfo {author} {\bibfnamefont {B.~C.}\ \bibnamefont
  {Johnson}}\ and\ \bibinfo {author} {\bibfnamefont {C.~S.}\ \bibnamefont
  {Campbell}},\ }\bibfield  {title} {\bibinfo {title} {Drop height and volume
  control the mobility of long-runout landslides on the earth and mars},\
  }\href@noop {} {\bibfield  {journal} {\bibinfo  {journal} {Geophys. Res.
  Lett.}\ }\textbf {\bibinfo {volume} {44}},\ \bibinfo {pages} {12} (\bibinfo
  {year} {2017})}\BibitemShut {NoStop}%
\bibitem [{\citenamefont {Heim}(1932)}]{heim1932bergsturz}%
  \BibitemOpen
  \bibfield  {author} {\bibinfo {author} {\bibfnamefont {A.}~\bibnamefont
  {Heim}},\ }\href@noop {} {\emph {\bibinfo {title} {Bergsturz und
  menschenleben}}},\ \bibinfo {number} {20}\ (\bibinfo  {publisher} {Fretz \&
  Wasmuth},\ \bibinfo {year} {1932})\BibitemShut {NoStop}%
\bibitem [{see()}]{seeSM}%
  \BibitemOpen
  \href@noop {} {}\bibinfo {note} {See Supplemental Material at [URL will be
  inserted by publisher], which includes Refs. [?-?] for additional details of
  the experiments and analysis.}\BibitemShut {Stop}%
\bibitem [{\citenamefont {Jop}\ \emph {et~al.}(2006)\citenamefont {Jop},
  \citenamefont {Forterre},\ and\ \citenamefont
  {Pouliquen}}]{jop2006constitutive}%
  \BibitemOpen
  \bibfield  {author} {\bibinfo {author} {\bibfnamefont {P.}~\bibnamefont
  {Jop}}, \bibinfo {author} {\bibfnamefont {Y.}~\bibnamefont {Forterre}},\ and\
  \bibinfo {author} {\bibfnamefont {O.}~\bibnamefont {Pouliquen}},\ }\bibfield
  {title} {\bibinfo {title} {A constitutive law for dense granular flows},\
  }\href@noop {} {\bibfield  {journal} {\bibinfo  {journal} {Nature}\ }\textbf
  {\bibinfo {volume} {441}},\ \bibinfo {pages} {727} (\bibinfo {year}
  {2006})}\BibitemShut {NoStop}%
\bibitem [{\citenamefont {Forterre}\ and\ \citenamefont
  {Pouliquen}(2008)}]{forterre2008flows}%
  \BibitemOpen
  \bibfield  {author} {\bibinfo {author} {\bibfnamefont {Y.}~\bibnamefont
  {Forterre}}\ and\ \bibinfo {author} {\bibfnamefont {O.}~\bibnamefont
  {Pouliquen}},\ }\bibfield  {title} {\bibinfo {title} {Flows of dense granular
  media},\ }\href@noop {} {\bibfield  {journal} {\bibinfo  {journal} {Ann. Rev.
  Fl. Mech.}\ }\textbf {\bibinfo {volume} {40}},\ \bibinfo {pages} {1}
  (\bibinfo {year} {2008})}\BibitemShut {NoStop}%
\bibitem [{\citenamefont {Lajeunesse}\ \emph {et~al.}(2005)\citenamefont
  {Lajeunesse}, \citenamefont {Monnier},\ and\ \citenamefont
  {Homsy}}]{lajeunesse2005granular}%
  \BibitemOpen
  \bibfield  {author} {\bibinfo {author} {\bibfnamefont {E.}~\bibnamefont
  {Lajeunesse}}, \bibinfo {author} {\bibfnamefont {J.}~\bibnamefont
  {Monnier}},\ and\ \bibinfo {author} {\bibfnamefont {G.}~\bibnamefont
  {Homsy}},\ }\bibfield  {title} {\bibinfo {title} {Granular slumping on a
  horizontal surface},\ }\href@noop {} {\bibfield  {journal} {\bibinfo
  {journal} {Phys. Fl.}\ }\textbf {\bibinfo {volume} {17}},\ \bibinfo {pages}
  {103302} (\bibinfo {year} {2005})}\BibitemShut {NoStop}%
\bibitem [{\citenamefont {Pirulli}\ and\ \citenamefont
  {Mangeney}(2008)}]{pirulli2008results}%
  \BibitemOpen
  \bibfield  {author} {\bibinfo {author} {\bibfnamefont {M.}~\bibnamefont
  {Pirulli}}\ and\ \bibinfo {author} {\bibfnamefont {A.}~\bibnamefont
  {Mangeney}},\ }\bibfield  {title} {\bibinfo {title} {Results of back-analysis
  of the propagation of rock avalanches as a function of the assumed
  rheology},\ }\href@noop {} {\bibfield  {journal} {\bibinfo  {journal} {Rock
  Mechanics and Rock Engineering}\ }\textbf {\bibinfo {volume} {41}},\ \bibinfo
  {pages} {59} (\bibinfo {year} {2008})}\BibitemShut {NoStop}%
\bibitem [{\citenamefont {Parez}\ and\ \citenamefont
  {Aharonov}(2015)}]{parez2015long}%
  \BibitemOpen
  \bibfield  {author} {\bibinfo {author} {\bibfnamefont {S.}~\bibnamefont
  {Parez}}\ and\ \bibinfo {author} {\bibfnamefont {E.}~\bibnamefont
  {Aharonov}},\ }\bibfield  {title} {\bibinfo {title} {Long runout landslides:
  a solution from granular mechanics},\ }\href@noop {} {\bibfield  {journal}
  {\bibinfo  {journal} {Front. Phys.}\ }\textbf {\bibinfo {volume} {3}},\
  \bibinfo {pages} {80} (\bibinfo {year} {2015})}\BibitemShut {NoStop}%
\bibitem [{\citenamefont {Kesseler}\ \emph {et~al.}(2020)\citenamefont
  {Kesseler}, \citenamefont {Heller},\ and\ \citenamefont
  {Turnbull}}]{kesseler2020grain}%
  \BibitemOpen
  \bibfield  {author} {\bibinfo {author} {\bibfnamefont {M.}~\bibnamefont
  {Kesseler}}, \bibinfo {author} {\bibfnamefont {V.}~\bibnamefont {Heller}},\
  and\ \bibinfo {author} {\bibfnamefont {B.}~\bibnamefont {Turnbull}},\
  }\bibfield  {title} {\bibinfo {title} {Grain reynolds number scale effects in
  dry granular slides},\ }\href@noop {} {\bibfield  {journal} {\bibinfo
  {journal} {Journal of Geophysical Research: Earth Surface}\ }\textbf
  {\bibinfo {volume} {125}},\ \bibinfo {pages} {e2019JF005347} (\bibinfo {year}
  {2020})}\BibitemShut {NoStop}%
\bibitem [{\citenamefont {Fei}\ \emph {et~al.}(2020)\citenamefont {Fei},
  \citenamefont {Jie}, \citenamefont {Sun},\ and\ \citenamefont
  {Chen}}]{fei2020particle}%
  \BibitemOpen
  \bibfield  {author} {\bibinfo {author} {\bibfnamefont {J.}~\bibnamefont
  {Fei}}, \bibinfo {author} {\bibfnamefont {Y.}~\bibnamefont {Jie}}, \bibinfo
  {author} {\bibfnamefont {X.}~\bibnamefont {Sun}},\ and\ \bibinfo {author}
  {\bibfnamefont {X.}~\bibnamefont {Chen}},\ }\bibfield  {title} {\bibinfo
  {title} {Particle size effects on small-scale avalanches and a $\mu$ (i)
  rheology-based simulation},\ }\href@noop {} {\bibfield  {journal} {\bibinfo
  {journal} {Computers and Geotechnics}\ }\textbf {\bibinfo {volume} {126}},\
  \bibinfo {pages} {103737} (\bibinfo {year} {2020})}\BibitemShut {NoStop}%
\bibitem [{\citenamefont {Johnson}\ \emph {et~al.}(2016)\citenamefont
  {Johnson}, \citenamefont {Campbell},\ and\ \citenamefont
  {Melosh}}]{johnson2016reduction}%
  \BibitemOpen
  \bibfield  {author} {\bibinfo {author} {\bibfnamefont {B.~C.}\ \bibnamefont
  {Johnson}}, \bibinfo {author} {\bibfnamefont {C.~S.}\ \bibnamefont
  {Campbell}},\ and\ \bibinfo {author} {\bibfnamefont {H.~J.}\ \bibnamefont
  {Melosh}},\ }\bibfield  {title} {\bibinfo {title} {The reduction of friction
  in long runout landslides as an emergent phenomenon},\ }\href@noop {}
  {\bibfield  {journal} {\bibinfo  {journal} {J. Geophys. Res. Earth Surface}\
  }\textbf {\bibinfo {volume} {121}},\ \bibinfo {pages} {881} (\bibinfo {year}
  {2016})}\BibitemShut {NoStop}%
\bibitem [{\citenamefont {Cerbus}\ \emph {et~al.}(2024)\citenamefont {Cerbus},
  \citenamefont {Brivady}, \citenamefont {Faug},\ and\ \citenamefont
  {Kellay}}]{companionPRE}%
  \BibitemOpen
  \bibfield  {author} {\bibinfo {author} {\bibfnamefont {R.~T.}\ \bibnamefont
  {Cerbus}}, \bibinfo {author} {\bibfnamefont {L.}~\bibnamefont {Brivady}},
  \bibinfo {author} {\bibfnamefont {T.}~\bibnamefont {Faug}},\ and\ \bibinfo
  {author} {\bibfnamefont {H.}~\bibnamefont {Kellay}},\ }\bibfield  {title}
  {\bibinfo {title} {Air drag controls the runout of small laboratory
  landslides},\ }\href {https://doi.org/10.1103/PhysRevE.109.064907} {\bibfield
   {journal} {\bibinfo  {journal} {Phys. Rev. E}\ }\textbf {\bibinfo {volume}
  {109}},\ \bibinfo {pages} {064907} (\bibinfo {year} {2024})},\ \bibinfo
  {note} {companion paper}\BibitemShut {NoStop}%
\bibitem [{\citenamefont {Lucas}\ \emph {et~al.}(2011)\citenamefont {Lucas},
  \citenamefont {Mangeney}, \citenamefont {M{\`e}ge},\ and\ \citenamefont
  {Bouchut}}]{lucas2011influence}%
  \BibitemOpen
  \bibfield  {author} {\bibinfo {author} {\bibfnamefont {A.}~\bibnamefont
  {Lucas}}, \bibinfo {author} {\bibfnamefont {A.}~\bibnamefont {Mangeney}},
  \bibinfo {author} {\bibfnamefont {D.}~\bibnamefont {M{\`e}ge}},\ and\
  \bibinfo {author} {\bibfnamefont {F.}~\bibnamefont {Bouchut}},\ }\bibfield
  {title} {\bibinfo {title} {Influence of the scar geometry on landslide
  dynamics and deposits: Application to martian landslides},\ }\href@noop {}
  {\bibfield  {journal} {\bibinfo  {journal} {Journal of Geophysical Research:
  Planets}\ }\textbf {\bibinfo {volume} {116}} (\bibinfo {year}
  {2011})}\BibitemShut {NoStop}%
\bibitem [{\citenamefont {McElwaine}\ and\ \citenamefont
  {Nishimura}(2001)}]{mcelwaine2001ping}%
  \BibitemOpen
  \bibfield  {author} {\bibinfo {author} {\bibfnamefont {J.}~\bibnamefont
  {McElwaine}}\ and\ \bibinfo {author} {\bibfnamefont {K.}~\bibnamefont
  {Nishimura}},\ }\bibfield  {title} {\bibinfo {title} {Ping-pong ball
  avalanche experiments},\ }\href@noop {} {\bibfield  {journal} {\bibinfo
  {journal} {Ann. Glaciol.}\ }\textbf {\bibinfo {volume} {32}},\ \bibinfo
  {pages} {241} (\bibinfo {year} {2001})}\BibitemShut {NoStop}%
\bibitem [{\citenamefont {Nishimura}\ \emph {et~al.}(1998)\citenamefont
  {Nishimura}, \citenamefont {Keller}, \citenamefont {McElwaine},\ and\
  \citenamefont {Nohguchi}}]{nishimura1998ping}%
  \BibitemOpen
  \bibfield  {author} {\bibinfo {author} {\bibfnamefont {K.}~\bibnamefont
  {Nishimura}}, \bibinfo {author} {\bibfnamefont {S.}~\bibnamefont {Keller}},
  \bibinfo {author} {\bibfnamefont {J.}~\bibnamefont {McElwaine}},\ and\
  \bibinfo {author} {\bibfnamefont {Y.}~\bibnamefont {Nohguchi}},\ }\bibfield
  {title} {\bibinfo {title} {Ping-pong ball avalanche at a ski jump},\
  }\href@noop {} {\bibfield  {journal} {\bibinfo  {journal} {Granular matter}\
  }\textbf {\bibinfo {volume} {1}},\ \bibinfo {pages} {51} (\bibinfo {year}
  {1998})}\BibitemShut {NoStop}%
\bibitem [{\citenamefont {Kosugi}\ \emph {et~al.}(1995)\citenamefont {Kosugi},
  \citenamefont {Sato}, \citenamefont {Abe}, \citenamefont {Nohguchi},
  \citenamefont {Yamada}, \citenamefont {Nishimura},\ and\ \citenamefont
  {Izumi}}]{kosugi1995table}%
  \BibitemOpen
  \bibfield  {author} {\bibinfo {author} {\bibfnamefont {K.}~\bibnamefont
  {Kosugi}}, \bibinfo {author} {\bibfnamefont {A.}~\bibnamefont {Sato}},
  \bibinfo {author} {\bibfnamefont {O.}~\bibnamefont {Abe}}, \bibinfo {author}
  {\bibfnamefont {Y.}~\bibnamefont {Nohguchi}}, \bibinfo {author}
  {\bibfnamefont {Y.}~\bibnamefont {Yamada}}, \bibinfo {author} {\bibfnamefont
  {K.}~\bibnamefont {Nishimura}},\ and\ \bibinfo {author} {\bibfnamefont
  {K.}~\bibnamefont {Izumi}},\ }\bibfield  {title} {\bibinfo {title} {Table
  tennis ball avalanche experiments},\ }in\ \href@noop {} {\emph {\bibinfo
  {booktitle} {ISSW94 Proceedings, International Snow science Workshop, October
  30-November}}},\ Vol.~\bibinfo {volume} {3}\ (\bibinfo {year} {1995})\ pp.\
  \bibinfo {pages} {636--642}\BibitemShut {NoStop}%
\bibitem [{\citenamefont {Cerbus}\ \emph {et~al.}(2021)\citenamefont {Cerbus},
  \citenamefont {Faug},\ and\ \citenamefont {Kellay}}]{cerbus2021landslide}%
  \BibitemOpen
  \bibfield  {author} {\bibinfo {author} {\bibfnamefont {R.~T.}\ \bibnamefont
  {Cerbus}}, \bibinfo {author} {\bibfnamefont {T.}~\bibnamefont {Faug}},\ and\
  \bibinfo {author} {\bibfnamefont {H.}~\bibnamefont {Kellay}},\ }\bibfield
  {title} {\bibinfo {title} {A landslide granular phase transition},\ }in\
  \href@noop {} {\emph {\bibinfo {booktitle} {EPJ Web of Conferences}}},\ Vol.\
  \bibinfo {volume} {249}\ (\bibinfo {organization} {EDP Sciences},\ \bibinfo
  {year} {2021})\ p.\ \bibinfo {pages} {03003}\BibitemShut {NoStop}%
\bibitem [{\citenamefont {Moro}\ \emph {et~al.}(2010)\citenamefont {Moro},
  \citenamefont {Faug}, \citenamefont {Bellot},\ and\ \citenamefont
  {Ousset}}]{moro2010large}%
  \BibitemOpen
  \bibfield  {author} {\bibinfo {author} {\bibfnamefont {F.}~\bibnamefont
  {Moro}}, \bibinfo {author} {\bibfnamefont {T.}~\bibnamefont {Faug}}, \bibinfo
  {author} {\bibfnamefont {H.}~\bibnamefont {Bellot}},\ and\ \bibinfo {author}
  {\bibfnamefont {F.}~\bibnamefont {Ousset}},\ }\bibfield  {title} {\bibinfo
  {title} {Large mobility of dry snow avalanches: Insights from small-scale
  laboratory tests on granular avalanches of bidisperse materials},\ }\href
  {https://doi.org/https://doi.org/10.1016/j.coldregions.2010.02.011}
  {\bibfield  {journal} {\bibinfo  {journal} {Cold Regions Science and
  Technology}\ }\textbf {\bibinfo {volume} {62}},\ \bibinfo {pages} {55}
  (\bibinfo {year} {2010})}\BibitemShut {NoStop}%
\bibitem [{\citenamefont {Linares-Guerrero}\ \emph {et~al.}(2007)\citenamefont
  {Linares-Guerrero}, \citenamefont {Goujon},\ and\ \citenamefont
  {Zenit}}]{linares2007increased}%
  \BibitemOpen
  \bibfield  {author} {\bibinfo {author} {\bibfnamefont {E.}~\bibnamefont
  {Linares-Guerrero}}, \bibinfo {author} {\bibfnamefont {C.}~\bibnamefont
  {Goujon}},\ and\ \bibinfo {author} {\bibfnamefont {R.}~\bibnamefont
  {Zenit}},\ }\bibfield  {title} {\bibinfo {title} {Increased mobility of
  bidisperse granular avalanches},\ }\href@noop {} {\bibfield  {journal}
  {\bibinfo  {journal} {Journal of Fluid Mechanics}\ }\textbf {\bibinfo
  {volume} {593}},\ \bibinfo {pages} {475} (\bibinfo {year}
  {2007})}\BibitemShut {NoStop}%
\bibitem [{\citenamefont {Lai}\ \emph {et~al.}(2017)\citenamefont {Lai},
  \citenamefont {Vallejo}, \citenamefont {Zhou}, \citenamefont {Ma},
  \citenamefont {Espitia}, \citenamefont {Caicedo},\ and\ \citenamefont
  {Chang}}]{lai2017collapse}%
  \BibitemOpen
  \bibfield  {author} {\bibinfo {author} {\bibfnamefont {Z.}~\bibnamefont
  {Lai}}, \bibinfo {author} {\bibfnamefont {L.~E.}\ \bibnamefont {Vallejo}},
  \bibinfo {author} {\bibfnamefont {W.}~\bibnamefont {Zhou}}, \bibinfo {author}
  {\bibfnamefont {G.}~\bibnamefont {Ma}}, \bibinfo {author} {\bibfnamefont
  {J.~M.}\ \bibnamefont {Espitia}}, \bibinfo {author} {\bibfnamefont
  {B.}~\bibnamefont {Caicedo}},\ and\ \bibinfo {author} {\bibfnamefont
  {X.}~\bibnamefont {Chang}},\ }\bibfield  {title} {\bibinfo {title} {Collapse
  of granular columns with fractal particle size distribution: Implications for
  understanding the role of small particles in granular flows},\ }\href@noop {}
  {\bibfield  {journal} {\bibinfo  {journal} {Geophysical Research Letters}\
  }\textbf {\bibinfo {volume} {44}},\ \bibinfo {pages} {12} (\bibinfo {year}
  {2017})}\BibitemShut {NoStop}%
\bibitem [{\citenamefont {Pong{\'o}}\ \emph {et~al.}(2021)\citenamefont
  {Pong{\'o}}, \citenamefont {Stiga}, \citenamefont {T{\"o}r{\"o}k},
  \citenamefont {L{\'e}vay}, \citenamefont {Szab{\'o}}, \citenamefont
  {Stannarius}, \citenamefont {Hidalgo},\ and\ \citenamefont
  {B{\"o}rzs{\"o}nyi}}]{pongo2021flow}%
  \BibitemOpen
  \bibfield  {author} {\bibinfo {author} {\bibfnamefont {T.}~\bibnamefont
  {Pong{\'o}}}, \bibinfo {author} {\bibfnamefont {V.}~\bibnamefont {Stiga}},
  \bibinfo {author} {\bibfnamefont {J.}~\bibnamefont {T{\"o}r{\"o}k}}, \bibinfo
  {author} {\bibfnamefont {S.}~\bibnamefont {L{\'e}vay}}, \bibinfo {author}
  {\bibfnamefont {B.}~\bibnamefont {Szab{\'o}}}, \bibinfo {author}
  {\bibfnamefont {R.}~\bibnamefont {Stannarius}}, \bibinfo {author}
  {\bibfnamefont {R.~C.}\ \bibnamefont {Hidalgo}},\ and\ \bibinfo {author}
  {\bibfnamefont {T.}~\bibnamefont {B{\"o}rzs{\"o}nyi}},\ }\bibfield  {title}
  {\bibinfo {title} {Flow in an hourglass: particle friction and stiffness
  matter},\ }\href@noop {} {\bibfield  {journal} {\bibinfo  {journal} {New
  Journal of Physics}\ }\textbf {\bibinfo {volume} {23}},\ \bibinfo {pages}
  {023001} (\bibinfo {year} {2021})}\BibitemShut {NoStop}%
\bibitem [{\citenamefont {Inaba}\ \emph {et~al.}(2017)\citenamefont {Inaba},
  \citenamefont {Tamaki}, \citenamefont {Ikebukuro}, \citenamefont {Yamada},
  \citenamefont {Ozaki},\ and\ \citenamefont {Yoshida}}]{inaba2017effect}%
  \BibitemOpen
  \bibfield  {author} {\bibinfo {author} {\bibfnamefont {Y.}~\bibnamefont
  {Inaba}}, \bibinfo {author} {\bibfnamefont {S.}~\bibnamefont {Tamaki}},
  \bibinfo {author} {\bibfnamefont {H.}~\bibnamefont {Ikebukuro}}, \bibinfo
  {author} {\bibfnamefont {K.}~\bibnamefont {Yamada}}, \bibinfo {author}
  {\bibfnamefont {H.}~\bibnamefont {Ozaki}},\ and\ \bibinfo {author}
  {\bibfnamefont {K.}~\bibnamefont {Yoshida}},\ }\bibfield  {title} {\bibinfo
  {title} {Effect of changing table tennis ball material from celluloid to
  plastic on the post-collision ball trajectory},\ }\href@noop {} {\bibfield
  {journal} {\bibinfo  {journal} {Journal of Human Kinetics}\ }\textbf
  {\bibinfo {volume} {55}},\ \bibinfo {pages} {29} (\bibinfo {year}
  {2017})}\BibitemShut {NoStop}%
\bibitem [{\citenamefont {Aaron}\ and\ \citenamefont
  {McDougall}(2019)}]{aaron2019rock}%
  \BibitemOpen
  \bibfield  {author} {\bibinfo {author} {\bibfnamefont {J.}~\bibnamefont
  {Aaron}}\ and\ \bibinfo {author} {\bibfnamefont {S.}~\bibnamefont
  {McDougall}},\ }\bibfield  {title} {\bibinfo {title} {Rock avalanche
  mobility: The role of path material},\ }\href@noop {} {\bibfield  {journal}
  {\bibinfo  {journal} {Eng. Geol.}\ }\textbf {\bibinfo {volume} {257}},\
  \bibinfo {pages} {105126} (\bibinfo {year} {2019})}\BibitemShut {NoStop}%
\bibitem [{\citenamefont {McSaveney}(2002)}]{mcsaveney2002recent}%
  \BibitemOpen
  \bibfield  {author} {\bibinfo {author} {\bibfnamefont {M.}~\bibnamefont
  {McSaveney}},\ }\bibfield  {title} {\bibinfo {title} {Recent rockfalls and
  rock avalanches in mount cook national park, new zealand},\ }in\ \href@noop
  {} {\emph {\bibinfo {booktitle} {Catastrophic Landslides: Effects,
  Occurrence, and Mechanisms}}},\ Vol.~\bibinfo {volume} {15}\ (\bibinfo
  {publisher} {Reviews in Engineering Geology},\ \bibinfo {year} {2002})\ pp.\
  \bibinfo {pages} {35--70}\BibitemShut {NoStop}%
\bibitem [{\citenamefont {Pfiffner}(2022)}]{pfiffner2022flims}%
  \BibitemOpen
  \bibfield  {author} {\bibinfo {author} {\bibfnamefont {O.~A.}\ \bibnamefont
  {Pfiffner}},\ }\bibfield  {title} {\bibinfo {title} {The flims rock
  avalanche: structure and consequences},\ }\href@noop {} {\bibfield  {journal}
  {\bibinfo  {journal} {Swiss Journal of Geosciences}\ }\textbf {\bibinfo
  {volume} {115}},\ \bibinfo {pages} {1} (\bibinfo {year} {2022})}\BibitemShut
  {NoStop}%
\bibitem [{\citenamefont {Cardinali}\ \emph {et~al.}(2002)\citenamefont
  {Cardinali}, \citenamefont {Reichenbach}, \citenamefont {Guzzetti},
  \citenamefont {Ardizzone}, \citenamefont {Antonini}, \citenamefont {Galli},
  \citenamefont {Cacciano}, \citenamefont {Castellani},\ and\ \citenamefont
  {Salvati}}]{cardinali2002geomorphological}%
  \BibitemOpen
  \bibfield  {author} {\bibinfo {author} {\bibfnamefont {M.}~\bibnamefont
  {Cardinali}}, \bibinfo {author} {\bibfnamefont {P.}~\bibnamefont
  {Reichenbach}}, \bibinfo {author} {\bibfnamefont {F.}~\bibnamefont
  {Guzzetti}}, \bibinfo {author} {\bibfnamefont {F.}~\bibnamefont {Ardizzone}},
  \bibinfo {author} {\bibfnamefont {G.}~\bibnamefont {Antonini}}, \bibinfo
  {author} {\bibfnamefont {M.}~\bibnamefont {Galli}}, \bibinfo {author}
  {\bibfnamefont {M.}~\bibnamefont {Cacciano}}, \bibinfo {author}
  {\bibfnamefont {M.}~\bibnamefont {Castellani}},\ and\ \bibinfo {author}
  {\bibfnamefont {P.}~\bibnamefont {Salvati}},\ }\bibfield  {title} {\bibinfo
  {title} {A geomorphological approach to the estimation of landslide hazards
  and risks in umbria, central italy},\ }\href@noop {} {\bibfield  {journal}
  {\bibinfo  {journal} {Nat. Hazards Earth Syst. Sci.}\ }\textbf {\bibinfo
  {volume} {2}},\ \bibinfo {pages} {57} (\bibinfo {year} {2002})}\BibitemShut
  {NoStop}%
\end{thebibliography}%

\end{document}